\documentstyle[prb,aps,psfig,twocolumn,preprint]{revtex}
\begin{document}

\onecolumn

\title{Implementation of 
analytical Hartree-Fock gradients for periodic systems}
\author{K. Doll} 
\address{CLRC, Daresbury Laboratory, Daresbury, Warrington, WA4 4AD, UK}
\address{Institut f\"ur Mathematische Physik, TU Braunschweig,
Mendelssohnstra{\ss}e 3, D-38106 Braunschweig}

\maketitle

\begin{abstract}
We describe the implementation of
 analytical Hartree-Fock gradients for periodic
systems in the code CRYSTAL, emphasizing the technical aspects
of this task. The code is now capable of calculating analytical
derivatives with respect to nuclear coordinates
for systems periodic in 0, 1, 2 and 3 dimensions
(i.e. molecules, polymers, slabs and solids). Both closed-shell restricted and
unrestricted Hartree-Fock gradients have been implemented.
 A comparison with
numerical derivatives shows that the forces are highly accurate.

\end{abstract}

\pacs{ }

\narrowtext
\section{Introduction}

The determination of equilibrium structure is one of the most important
targets in electronic structure calculations. In surface
science especially, theoretical calculations of surface structures
are of high importance to explain and support experimental
results. Therefore, a fast structural optimization is an important
issue in modern electronic structure codes. Finding minima in
energy surfaces is substantially simplified by the availability of
analytical gradients. As a rule of thumb, availability of analytical gradients
improves the efficiency by  a factor of order $N$ with $N$
being the number of parameters to be optimized. UK's Collaborative
Computational Project 3 has therefore supported the implementation
of analytical gradients in the electronic structure code CRYSTAL
\cite{CRYSTALbuch,Pisani1980,Manual,Vicbook}.
This implementation will also be valuable for future projects which
require  analytical gradients as a prerequisite. Another advantage
of having analytical gradients is that higher derivatives can be obtained
with less numerical noise (e.g. the
2nd derivative has less numerical noise when only one numerical 
differentiation is necessary).

CRYSTAL is capable of performing Hartree-Fock and density-functional
calculations for systems with any periodicity (i.e. molecules,
polymers, slabs and solids). The periodicity is "cleanly" implemented
in the sense that, for example, a slab is considered as an object
periodic in two dimensions and is {\em not} repeated in the third dimension
with one slab being separated from the others by vacuum layers.
The code is based on Gaussian type
orbitals and the technology is therefore in many parts similar
to that of molecular quantum chemistry codes.  As the
density-functional part of the code relies in big parts on the
Hartree-Fock part, the strategy of the
project was to implement Hartree-Fock gradients first.

The implementation of Hartree-Fock gradients for multicenter basis sets
was pioneered by Pulay\cite{Pulay};
the theory had already been derived earlier independently \cite{Bratoz}.
Meanwhile, analytical gradients have been implemented in many molecular
codes, and several review articles have appeared (see, e.g., references
\onlinecite{PulayAdv,PulayChapter,Helgaker,HelgakerJorgensen1992,SchlegelYarkony,PulayYarkony,Schlegel2000}).

Substantial work has also been done in the case of one-dimensional
periodicity:
Hartree-Fock gradients with respect to nuclear coordinates and
with respect to the lattice vector have already been implemented in codes
periodic in one dimension \cite{Teramae,HirataIwata,Jacquemin}. Moreover,
correlated
calculations based on the MP2 scheme \cite{Suhai,SunBartlett}
and MP2 gradients \cite{HirataIwata} have been coded. 
Also, density functional gradients
have been implemented \cite{HirataIwataDFT,KudinScuseria}.
Even second derivatives at the Hartree-Fock level
have meanwhile been coded \cite{HirataIwata2nd}.

The implementation of Hartree-Fock gradients with respect to
nuclear coordinates in CRYSTAL is to the best of our
knowledge the first implementation for the case of 2- and 3-dimensional
periodicity.
The aim of this article is to  describe the implementation of the
gradients in the code, with an emphasis on the technical aspects.
Therefore, the article is supposed to complement our first article
on the purely theoretical aspects \cite{IJQC}. An attempt of
a detailed description is made; however, as the whole code is undergoing
constant changes, it can not be too detailed. For example, it did
not seem advisable to give any variable names because they have already
undergone major changes after the code moved to Fortran 90 with the
possibility of longer variable names. 

The article is structured as follows: In section \ref{BasisHFsection}, we give
a brief introduction to Gaussian and Hermite Gaussian type basis functions.
The definition of the density matrix is given in section \ref{DMsection}.
The individual integrals, their derivatives, and details of the
implementation
are discussed in section \ref{Integralsection}. Formulas for total
energy and gradient are given in section \ref{Calculationofderivatives}.
The structure of the gradient code is explained 
in section \ref{Codestructure},
followed by examples in section \ref{Examplesection} and the conclusion.

\section{Basis functions}

\label{BasisHFsection}

Two sets of basis functions are relevant for CRYSTAL:
firstly, unnormalized spherical Gaussian type functions,
in a polar coordinate system 
characterized by the set of variables $(|\vec r|,\vartheta,\varphi)$,
and centered at $\vec A$.
They are defined as

\begin{equation}
S(\alpha,\vec r-\vec A,n,l,m)={|\vec r-\vec A|}^{2n+l} 
 {\rm P}_l^{|m|}(\cos \ \vartheta)
\exp({\rm i}m\varphi)\exp (-\alpha {|\vec r-\vec A|}^2)
\end{equation}

with ${\rm P}_l^{|m|}$ being the associated Legendre function.
CRYSTAL uses real spherical Gaussian type functions 
defined as

\begin{eqnarray*}
R(\alpha,\vec r-\vec A,n,l,0)=S(\alpha,\vec r-\vec A,n,l,0)\\
R(\alpha,\vec r-\vec A,n,l,|m|)={\rm Re \ } S(\alpha,\vec r-\vec A,n,l,|m|)\\
R(\alpha,\vec r-\vec A,n,l,-|m|)={\rm Im \ } S(\alpha,\vec r-\vec A,n,l,|m|)
\end{eqnarray*}

This is in the following denoted as 
$\phi_{\mu}(\alpha,\vec  r - \vec A_{\mu},n,l,m)=
N_{\mu} R(\alpha,\vec r - \vec A_{\mu},n,l,m)$, 
with the normalization $N_{\mu}$. $\mu$ is an index enumerating
the basis functions
in the reference cell (e.g. the primitive unit cell).
In fact, CRYSTAL uses only basis functions with quantum number $n=0$ and
angular momentum $l$=0,1 or 2 (i.e. $s$, $p$ or $d$ functions). 

The exponents are defined by the user of the code. A huge amount
of basis sets for molecular calculations is
available in the literature and on the world wide web; also for
periodic systems a large number of basis sets has been published.
Molecular basis sets can, with a little effort, be adopted for 
solid state calculations. High exponents which are used to describe
core electrons do not have to be adjusted, but exponents with low
values (e.g. less than 1 $a_0^{-2}$, with $a_0$ being the Bohr radius) should
be reoptimized for the corresponding solid.
Very diffuse exponents should be omitted
because they cause linear dependence problems in periodic systems.

A second type of basis functions, which CRYSTAL uses internally
to evaluate the
integrals, is the
Hermite Gaussian type function (HGTF) which is defined as:

\begin{eqnarray}
\Lambda(\gamma,\vec r-\vec A,t,u,v)=
\bigg(\frac{\partial}{\partial A_x}\bigg)^t
\bigg(\frac{\partial}{\partial A_y}\bigg)^u
\bigg(\frac{\partial}{\partial A_z}\bigg)^v {\exp}(-\gamma
|\vec r-\vec A|^2)
\end{eqnarray}

CRYSTAL uses the  McMurchie-Davidson algorithm to evaluate the integrals.
The basic idea of this algorithm is to map the product of two
spherical Gaussian type functions  on two centers onto a set of
Hermite Gaussian type functions at one center.

\begin{eqnarray}
\label{S1S2lambda}
S(\tilde \alpha,\vec r-\vec B,\tilde n,\tilde l,\tilde m)S(\alpha,\vec r-\vec A,n,l,m)
=\nonumber \\
\sum_{t,u,v} E(\tilde n,\tilde l,\tilde m,n,l,m,t,u,v)
\Lambda(\gamma,\vec r-\vec P,t,u,v)
\end{eqnarray}

with $\gamma=\alpha+\tilde \alpha$ and 
$\vec P=\frac{\alpha \vec A+\tilde \alpha\vec B}{\alpha+\tilde \alpha}$.

The starting point $E(0,0,0,0,0,0,0,0,0)=\exp(-\frac{\alpha\tilde \alpha}
{\alpha+\tilde \alpha}
|\vec B-\vec A|^2)$ is derived from the Gaussian product rule
\cite{Boys}:

\begin{eqnarray}
\exp(-\alpha |\vec r-\vec A|^2)\exp(-\tilde \alpha |\vec r-\vec B|^2)=
\exp\bigg(-\frac{\alpha\tilde \alpha}{\alpha+\tilde \alpha}|\vec B-\vec A|^2\bigg)
\exp\bigg(-(\alpha+\tilde \alpha)
\bigg|\vec r -\frac {\alpha\vec A + \tilde \alpha \vec B}{\alpha+\tilde \alpha}
\bigg|^2\bigg)
\end{eqnarray}

As indicated in section \ref{Integralsection}, all the integrals 
can be expressed with the help of the coefficients
$E(\tilde n,\tilde l,\tilde m,n,l,m,t,u,v)$
\cite{McMurchieDavidson,VicNATO,VicCoulomb,Dovesi1983}.
These
coefficients are generated by recursion relations 
\cite{McMurchieDavidson,VicNATO}. They are zero for the case 
$t+u+v>2n +2 \tilde n+l+\tilde l$ and for all negative values of $t,u$ or $v$.
CRYSTAL uses only basis functions with $n=0$. Therefore, there are 
$\frac{(l+\tilde l+1)(l+\tilde l+2)(l+\tilde l+3)}{3!}$ coefficients 
$E(0,\tilde l,\tilde m,0,l,m,t,u,v)$
for fixed values of $l,m,\tilde l,\tilde m$. As the maximum angular
quantum number is $l=2$, this results in  25 possible combinations 
of $m$ and $\tilde m$. Therefore, the maximum number of coefficients is
$25 \times 35=875$. These coefficients are 
pre-programmed in the subroutine DFAC3.
Pre-programming is the fastest possible way of evaluating these coefficients
which is important because this is one of the key issues of the integral
calculation. On the other hand, the code has become inflexible as no
$E$-coefficients are available for higher quantum numbers.

Derivatives of Gaussian type functions are again Gaussian type functions.
Therefore, the evaluation of gradients is closely related to the
evaluation of integrals. In a similar way as all the integrals
can be expressed with the help of coefficients $E$,
all the derivatives of the integrals
can be expressed with the help of coefficients for the gradients,
$G_x^{A},G_y^{A},G_z^A$. These $G$-coefficients can be obtained
with recursion relations derived by Saunders
\cite{Vicbook,IJQC}. The recursions are similar to the ones for the
$E$-coefficients.
However,
as the existing subroutine 
DFAC3 cannot compute the $G$-coefficients, the
recursions were newly coded. This has in addition the advantage that,
by small modifications of the new subroutines, 
$E$-coefficients for higher quantum numbers than $l=\tilde l=2$
can now be computed by recursion. There are three sets of $G$-coefficients
because of the three spatial directions. The $G$-coefficients are zero
for the case $t+u+v>2n +2 \tilde n+l+\tilde l+1$ 
and for all negative values of $t,u$ or $v$. This means that for 
a maximum quantum number of $l=2$, there are 
$3 \times 5 \times 5 \times 56=4200$ coefficients. Three other sets
of $G$-coefficients are necessary because of the second center.
However, the sets on the second center
 are closely related  to the sets on
the first center
and can be derived from them in an efficient way
\cite{HelgakerTaylor,Vicbook,IJQC}.

\section{Density matrix}
\label{DMsection}

After solving the Hartree-Fock equations \cite{Andre},
the crystalline orbitals are linear combinations of Bloch functions

\begin{equation}
\Psi_i(\vec r, \vec k)=\sum_{\mu} a_{\mu i}(\vec k)\psi_{\mu}(\vec r, \vec k)
\end{equation}

which are expanded in terms of real spherical Gaussian type functions

\begin{equation}
\psi_{\mu}(\vec r, \vec k)=N_{\mu} \sum_{\vec g}
R(\alpha,\vec r-\vec A_{\mu}-\vec g,n,l,m)
{\rm e}^{{\rm i} \vec k\vec g}
\end{equation}

The sum over $\vec g$ is over all direct lattice vectors.

In the case of closed shell, spin-restricted Hartree-Fock,
the spin-free density matrix in reciprocal space is defined as

\begin{eqnarray}
P_{\mu\nu}(\vec k)=2\sum_{i}a_{\mu i}(\vec k)a_{\nu i}^*(\vec k)
\Theta(\epsilon_F-\epsilon_i(\vec k))
\end{eqnarray}

with the Fermi energy $\epsilon_F$ and the Heaviside function $\Theta$; $i$
is an index enumerating the eigenvalues.

In the case of unrestricted Hartree-Fock (UHF) \cite{Apra}, 
we use the notation

\begin{equation}
\Psi_i^{\uparrow}(\vec r, \vec k)=\sum_{\mu} a^{\uparrow}_{\mu i}(\vec k)
\psi_{\mu}(\vec r, \vec k)
\end{equation}

and

\begin{equation}
\Psi_i^{\downarrow}(\vec r, \vec k)=\sum_{\mu} a^{\downarrow}_{\mu i}(\vec k)
\psi_{\mu}(\vec r, \vec k)
\end{equation}

for the crystalline orbitals with up and down spin, respectively. We define
the density matrices

\begin{eqnarray}
P_{\mu\nu}^{\uparrow}(\vec k)=\sum_{i}a_{\mu i}^{\uparrow}(\vec k)a_{\nu i}
^{* \uparrow}(\vec k)
\Theta(\epsilon_F-\epsilon_i^{\uparrow}(\vec k))
\end{eqnarray}

for up spin and 

\begin{eqnarray}
P_{\mu\nu}^{\downarrow}(\vec k)=\sum_{i}a_{\mu i}^{\downarrow}(\vec k)
a_{\nu i}
^{* \downarrow}(\vec k)
\Theta(\epsilon_F-\epsilon_i^{\downarrow}(\vec k))
\end{eqnarray}

for down spin. In the following, $P_{\mu\nu}$ refers to the sum
$P_{\mu\nu}^{\uparrow}+P_{\mu\nu}^{\downarrow}$ in the UHF case.

The density matrices in real space 
$P_{\mu\vec 0\nu\vec g}, P_{\mu\vec 0\nu\vec g}^{\uparrow}, 
P_{\mu\vec 0\nu\vec g}^{\downarrow}$
are obtained by Fourier transformation.

\section{Integrals and their derivatives}

\label{Integralsection}

The calculation of the integrals is fundamental to all quantum chemistry
programs. CRYSTAL uses two integral packages: a package derived from
GAUSSIAN70 \cite{GAUSSIAN70} is the default
for calculations when
only $s$ and $sp$ shells are used; alternatively Saunders' 
ATMOL Gaussian integral package can be used and it must be used for 
cases when $p$ or $d$ functions are involved. The implementation of gradients
has been done with routines based on the ATMOL package. This is not
a restriction, and it is possible to use routines based on GAUSSIAN70 
for the integrals and routines based on ATMOL
for the gradients.

The calculation of the
integrals is essentially controlled from MONMAD and MONIRR
for one-electron integrals and from SHELLC or SHELLX for the 
bielectronic integrals. SHELLC is used in the case of non-direct SCF,
i.e. when the integrals are written to disk and read in each cycle.
SHELLX is the direct version when the integrals are computed in each cycle
without storing them on disk. The direct mode is the preferred one when
the integral file is too big or when
input/output to disk is too slow. 
The gradients are computed only once after the last iteration, when
convergence is achieved. Therefore, a direct implementation of gradients
has been done. 

One of the bottlenecks of the CRYSTAL code is the restriction to
a highest quantum number of $l=2$, i.e. the code can only cope with
$s$, $p$, $sp$ and $d$ functions, but not with basis functions with
higher angular momentum. Introducing gradients, however, is similar
to increasing the quantum number from $d$ to $f$ for the corresponding
basis function. This means that many subroutines had to be extended
to higher quantum numbers, and array dimensions in the whole code
had to be adjusted.

\subsection{One-electron integrals}

In this section we summarize the appearing types of integrals and the
corresponding gradients. We restrict the description to the x-component
of the gradient; y- and z-component can be obtained in similar way.
Note that the integrals
 depend on the dimension because
of the Ewald scheme used.  Therefore, there are four different routines for
the one-electron integrals for the 
case of 0,1,2 and 3-dimensional periodicity: 
CJAT0, CJAT1, CJAT2  and CJAT3.
Similarly, four gradient routines have been developed which have been
given the preliminary names CJAT0G, CJAT1G, CJAT2G and CJAT3G.
These routines calculate all the one-electron integrals except for
the multipolar integrals which are computed in POLIPA (with the 
corresponding gradient routine OLIPAG).

\subsubsection{Overlap integral}

The basic integral is the overlap integral:

\begin{eqnarray} & &
S_{\mu\vec {g_1}\nu\vec {g_2}}= 
\int \phi_{\mu}
(\tilde \alpha,\vec r - \vec A_{\mu}-\vec g_1,\tilde n,\tilde l,\tilde m)
\phi_{\nu}(\alpha,\vec r - \vec A_{\nu}-\vec g_2,n,l,m){\rm d^3r}
= \nonumber \\  & & 
\int\sum_{t,u,v}
E(\tilde n,\tilde l,\tilde m,n,l,m,t,u,v)\Lambda(\gamma,\vec r-\vec P,t,u,v)
{\rm d^3r}=\\ \nonumber & &
E(\tilde n,\tilde l,\tilde m,n,l,m,0,0,0)
\left( \frac{\pi}{\gamma}\right)^{\frac{3}{2}}
\end{eqnarray}

The x-component of the gradient with respect to center $A_{\mu}$
is obtained as
\begin{eqnarray} & &
\frac{\partial}{\partial A_{\mu,x}}
S_{\mu\vec {g_1}\nu\vec {g_2}}=\\ & & \nonumber
\frac{\partial}{\partial A_{\mu,x}}
\int \phi_{\mu}
(\tilde \alpha,\vec r - \vec A_{\mu}-\vec g_1,\tilde n,\tilde l,\tilde m)
\phi_{\nu}(\alpha,\vec r - \vec A_{\nu}-\vec g_2,n,l,m){\rm d^3r}
= \nonumber \\ & &
\frac{\partial}{\partial A_{\mu,x}}
\int\sum_{t,u,v}
E(\tilde n,\tilde l,\tilde m,n,l,m,t,u,v)\Lambda(\gamma,\vec r-\vec P,t,u,v)
{\rm d^3r}= \nonumber \\ & &
\int\sum_{t,u,v}G_x^{A_\mu}
(\tilde n,\tilde l,\tilde m,n,l,m,t,u,v)\Lambda(\gamma,\vec r-\vec P,t,u,v)
{\rm d^3r}= \nonumber \\ & &
G_x^{A_{\mu}}
(\tilde n,\tilde l,\tilde m,n,l,m,0,0,0)
\left( \frac{\pi}{\gamma}\right)^{\frac{3}{2}}
\label{overlapgradienteqn}
\end{eqnarray}

Equation \ref{overlapgradienteqn} 
thus defines the coefficients $G_x^{A_{\mu}}$; similarly the coefficients
$G_y^{A_{\mu}}, G_z^{A_{\mu}}, G_x^{A_{\nu}}, G_y^{A_{\nu}}, G_z^{A_{\nu}}$ 
can be defined.

In the following, we use the identity

$S_{\mu\vec {g_1}\nu\vec {g_2}}=S_{\mu\vec {0}\nu(\vec {g_2}-\vec{g_1})}=
S_{\mu\vec {0}\nu\vec {g}}$.

\subsubsection{Kinetic energy integrals}

In equation \ref{kineticNATO}, the expression 
for the kinetic energy integrals for the case
of spherical Gaussian type functions is reiterated \cite{VicNATO}:

\begin{eqnarray} 
\label{kineticNATO}
& & 
T_{\mu\vec 0\nu\vec g}= \nonumber \\ & &
\int \phi_{\mu}(\tilde\alpha,\vec r-\vec A_{\mu},\tilde n,\tilde l,\tilde m
)\left(-\frac{1}{2}\Delta_{\vec r}\right)\phi_{\nu}(\alpha,
\vec r - \vec A_{\nu}-\vec g,n,l,m)
{\rm d^3r}
=  \nonumber \\ & & 
-n(2n+2l+1)
\int \phi_{\mu}(\tilde\alpha,\vec r - \vec A_{\mu},\tilde n,\tilde l,\tilde m)
\phi_{\nu}(\alpha,\vec r - \vec A_{\nu}-\vec g,n-1,l,m){\rm d^3r}
 +  \nonumber \\ & & 
\alpha(4n+2l+3)
\int \phi_{\mu}(\tilde\alpha,\vec r - \vec A_{\mu},\tilde n,\tilde l,\tilde m)
\phi_{\nu}(\alpha,\vec r - \vec A_{\nu}-\vec g,n,l,m){\rm d^3r}
- \nonumber \\ & & 
2\alpha^2 
\int \phi_{\mu}(\tilde\alpha,\vec r - \vec A_{\mu},\tilde n,\tilde l,\tilde m)
\phi_{\nu}(\alpha,\vec r - \vec A_{\nu}-\vec g,n+1,l,m){\rm d^3r}
= \nonumber \\ & & 
-n(2n+2l+1)\int\sum_{t,u,v}E(\tilde n,\tilde l,\tilde m,n-1,l,m,t,u,v)
\Lambda(\gamma,\vec r-\vec P,t,u,v){\rm d^3r}
+ \nonumber \\ & & 
\alpha(4n+2l+3)\int\sum_{t,u,v}E(\tilde n,\tilde l,\tilde m,n,l,m,t,u,v)
\Lambda(\gamma,\vec r-\vec P,t,u,v){\rm d^3r}
-\nonumber \\ & & 
2\alpha^2 \int\sum_{t,u,v}E(\tilde n,\tilde l,\tilde m,n+1,l,m,t,u,v)
\Lambda(\gamma,\vec r-\vec P,t,u,v){\rm d^3r}
\end{eqnarray}

The x-component of the gradient is therefore:

\begin{eqnarray} & & 
\frac{\partial}{\partial A_{\mu,x}}T_{\mu\vec 0\nu\vec g}= \nonumber \\ & & 
-n(2n+2l+1)\int\sum_{t,u,v}G_x^{A_{\mu}}
(\tilde n,\tilde l,\tilde m,n-1,l,m,t,u,v)\Lambda(\gamma,\vec r-\vec P,t,u,v)
{\rm d^3r}+
\nonumber \\ & & \alpha(4n+2l+3)\int\sum_{t,u,v}G_x^{A_{\mu}}
(\tilde n,\tilde l,\tilde m,n,l,m,t,u,v)\Lambda(\gamma,\vec r-\vec P,t,u,v)
{\rm d^3r}-
\nonumber \\ & & 2\alpha^2 \int\sum_{t,u,v}G_x^{A_{\mu}}
(\tilde n,\tilde l,\tilde m,n+1,l,m,t,u,v)\Lambda(\gamma,\vec r-\vec P,t,u,v)
{\rm d^3r}
\end{eqnarray}

As CRYSTAL uses spherical Gaussian type functions with $n=0$, this
reduces to

\begin{eqnarray} & &
\frac{\partial}{\partial A_{\mu,x}}T_{\mu\vec 0\nu\vec g}= \nonumber \\ & & 
\left( \frac{\pi}{\gamma}\right)^{\frac{3}{2}}
\alpha(4n+2l+3)G_x^{A_{\mu}}(0,\tilde l,\tilde m,0,l,m,0,0,0)- \nonumber \\ 
& & 2\left( \frac{\pi}{\gamma}\right)^{\frac{3}{2}}
\alpha^2 G_x^{A_{\mu}}(0,\tilde l,\tilde m,1,l,m,0,0,0)
\end{eqnarray}

Explicit differentiation with respect to the other center $\vec A_{\nu}$
is more difficult because the kinetic energy operator applies to that
center. However, the differentiation
can easily be avoided by applying translational
invariance:

\begin{eqnarray}
\frac{\partial}{\partial A_{\mu,x}}T_{\mu\vec 0\nu\vec g}=
-\frac{\partial}{\partial A_{\nu,x}}T_{\mu\vec 0\nu\vec g}
\end{eqnarray}

\subsubsection{Nuclear attraction integrals}

The nuclear attraction integrals are defined as

\begin{eqnarray} & & 
N_{\mu\vec 0\nu\vec g}= -\sum_a Z_a\int 
\phi_{\mu}(\tilde\alpha,\vec r-\vec A_{\mu},\tilde n,\tilde l,\tilde m)
A(\vec r-\vec A_{a})
\phi_{\nu}(\alpha,\vec r - \vec A_{\nu}-\vec g,n,l,m) {\rm d^3r}=
\nonumber \\ & & 
-\sum_a Z_a \int\sum_{t,u,v}E(\tilde n,\tilde l,\tilde m,n,l,m,t,u,v)
\Lambda(\gamma,\vec r-\vec P,t,u,v)
A{(\vec r-\vec A_{a})}{\rm d^3r}
\end{eqnarray}

where $A$ is the Coulomb potential function in the molecular case,
the Euler-MacLaurin potential function for systems periodic in one dimension
\cite{Vic1994}, Parry's potential function \cite{Parry} for
systems periodic in two dimensions, and the Ewald potential function for
systems periodic in three dimensions \cite{Ewald,Harris,VicCoulomb}.
The summation with respect to $a$ 
runs over all nuclei of the primitive unit cell.

The x-component of the
partial derivative with respect to the center $A_{\mu,x}$ is obtained as:

\begin{eqnarray} & & 
\frac{\partial}{\partial A_{\mu,x}}N_{\mu\vec 0\nu\vec g}=  \nonumber \\ & &
-\sum_a Z_a
\int\sum_{t,u,v}G_x^{A_{\mu}}(\tilde n,\tilde l,\tilde m,n,l,m,t,u,v)
\Lambda(\gamma,\vec r-\vec P,t,u,v)
A{(\vec r-\vec A_{a})}{\rm d^3r}
\end{eqnarray}

In the same way, the partial derivative with respect to $A_{\nu,x}$ 
is obtained.
The partial derivative
with respect to the set of third centers $\vec A_a$ is obtained
by translational invariance: for each center $\vec A_a$, there is a derivative 
with value \\
$-\frac{\partial}{\partial\vec 
A_{\mu}}-\frac{\partial}{\partial \vec A_{\nu}}$.

\subsubsection{Multipolar integrals}

The electronic charge density is expressed with a lattice basis as:

\begin{eqnarray}
\rho(\vec r)= -\sum_{\vec g,\mu ,\nu}P_{\nu\vec g\mu\vec 0}
\phi_{\mu}(\tilde\alpha,\vec r-\vec A_{\mu},\tilde n,\tilde l,\tilde m)
\phi_{\nu}(\alpha,\vec r - \vec A_{\nu}-\vec g,n,l,m)
\end{eqnarray}

Then, the Ewald potential due to this charge density is given by:

\begin{eqnarray}
\Phi^{ew}(\rho;\vec r)=
\int A(\vec r - \vec r')\rho(\vec r') d^3r'
\end{eqnarray}

The Ewald energy of the electons 
(i.e. the Ewald energy of the electrons in the primitive unit cell
with all the electrons) is obtained as

\begin{eqnarray}
E=\frac{1}{2}\int\int \rho(\vec r)A(\vec r - \vec r')\rho(\vec r') d^3rd^3r'
\end{eqnarray}

For efficiency reasons, the calculation of the Ewald potential is
done approximatively. A multipolar expansion up to
an order $L$ is performed for the charge distribution in the long range.
Therefore, the electrons do not feel the Ewald potential
created by the correct charge distribution, but the Ewald potential created
by the multipolar moments. It is thus necessary to compute the multipolar
moments of the charge distribution which are defined as

\begin{eqnarray} 
\eta_l^m(\rho_c;\vec A_c)=\int \rho_c(\vec r) 
X_l^m(\vec r-\vec A_{c}){\rm d^3r} 
\end{eqnarray}

with $X_l^m$ being regular solid harmonics \cite{VicCoulomb} and
the charge $\rho_c(\vec r)$ defined as

\begin{eqnarray} & &
\rho_c(\vec r)=-\sum_{\vec g,\mu \in c,\nu}P_{\nu\vec g\mu\vec 0}
\phi_{\mu}(\tilde\alpha,\vec r-\vec A_{\mu},\tilde n,\tilde l,\tilde m)
\phi_{\nu}(\alpha,\vec r - \vec A_{\nu}-\vec g,n,l,m)
= \nonumber \\ & &
-\sum_{\vec g,\mu \in c,\nu}P_{\nu\vec g\mu\vec 0} \sum_{t,u,v}
E(\tilde n,\tilde l,\tilde m,n,l,m,t,u,v)\Lambda(\gamma,\vec r-\vec P,t,u,v)
\end{eqnarray}

$c$ is an index for the shell. The total electronic charge $\rho(\vec r)$
is thus obtained by summing over all shells $c$:

\begin{eqnarray}
\rho(\vec r)=\sum_c \rho_c(\vec r)
\end{eqnarray}

In CRYSTAL,  the multipole is located at center $\vec A_{\mu}$ and
therefore it is convenient to take the derivative with respect
to center $\vec A_{\nu}$ and obtain the derivative with respect to 
$\vec A_{\mu}$ by translational invariance.
The expression computed for the gradients is thus

\begin{eqnarray} & &
-\sum_{\vec g,\mu \in c,\nu}P_{\nu\vec g\mu\vec 0}\int 
\frac{\partial}{\partial A_{\nu,x}}
\left(
\phi_{\mu}(\tilde\alpha,\vec r-\vec A_{\mu},\tilde n,\tilde l,\tilde m)
\phi_{\nu}(\alpha,\vec r - \vec A_{\nu}-\vec g,n,l,m)
X_l^m(\vec r-\vec A_{\mu})\right){\rm d^3r}
= \nonumber \\ & &
-\sum_{\vec g,\mu \in c,\nu}P_{\nu\vec g\mu\vec 0}\int \sum_{t,u,v} 
G_x^{A_{\nu}}(\tilde n,\tilde l,\tilde m,n,l,m,t,u,v)
\Lambda(\gamma,\vec r-\vec P,t,u,v)X_l^m(\vec r-\vec A_{\mu})
{\rm d^3r}
\end{eqnarray}

\subsubsection{Field integrals}

If the electronic charge distribution
is approximated with an expansion up to
the maximum quantum number $L$, the Ewald potential of this model charge
distribution is obtained as

\begin{eqnarray}
\Phi^{ew}(\rho^{model};\vec r)=\sum_c \Phi^{ew}(\rho_c^{model};\vec r)=
\sum_c \sum_{l=0}^L \sum_{m=-l}^l \eta_l^m(\rho_c;\vec A_c)
Z_l^m(\hat{\vec A_c})A(\vec r-\vec A_c)
\end{eqnarray}

with $Z_l^m(\hat {\vec A_c})$ being the spherical gradient 
operator in a renormalized form\cite{VicCoulomb}. 
The model charge distribution is expressed as

\begin{eqnarray}
\rho_c^{\rm model}(\vec r)
=\sum_{l=0}^{L} \sum_{m=-l}^{l} \eta_l^m(\rho_c;\vec A_c)
\delta_l^m(\vec A_c,\vec r)
\end{eqnarray}

and

\begin{eqnarray}
\delta_{l}^{m}(\vec {A_c},\vec r)=\lim_{\alpha \rightarrow \infty}
Z_{l}^{m}(\hat{\vec A_c})\Lambda(\alpha,\vec r-\vec A_c,0,0,0)
\end{eqnarray}

The integral of the electronic charge
distribution and the Ewald potential function is required which 
gives rise to the field integrals
which are defined as follows:

\begin{eqnarray} & &
M_{l\mu\vec 0\nu\vec gc}^m= \nonumber \\ & & Z_l^m(\hat{\vec A_c})\int
\phi_{\mu}(\tilde\alpha,\vec r-\vec A_{\mu},\tilde n,\tilde l,\tilde m)
\phi_{\nu}(\alpha,\vec r - \vec A_{\nu}-\vec g,n,l,m)
\bigg[A(\vec r-\vec A_c)-\sum_{\vec n}^{pen}
\frac{1}{|\vec r-\vec A_c-\vec n|}\bigg]{\rm d^3r}
=  \nonumber \\ & & Z_l^m(\hat{\vec A_c})\int\sum_{t,u,v}
E(\tilde n,\tilde l,\tilde m,n,l,m,t,u,v)\Lambda(\gamma,\vec r-\vec P,t,u,v)
\bigg[A(\vec r-\vec A_c)-\sum_{\vec n}^{pen}
\frac{1}{|\vec r-\vec A_c-\vec n|}\bigg]{\rm d^3r}
\end{eqnarray}

The term $\bigg[A(\vec r-\vec A_c)-\sum_{\vec n}^{pen}
\frac{1}{|\vec r-\vec A_c-\vec n|}\bigg]$ instead of  
$A(\vec r-\vec A_c)$ appears because the multipolar approximation
is only done for the charge distribution in the long range.
The penetration depth $pen$ is a certain threshold for which the
integrals are evaluated exactly \cite{VicCoulomb,Manual}.

For the gradients, the derivative with respect to all the centers is needed.
The partial derivative with respect to $A_{\mu,x}$ is obtained as

\begin{eqnarray} & &
\frac{\partial}{\partial A_{\mu,x}}M_{l\mu\vec 0\nu\vec gc}^m=
\nonumber \\ & & 
Z_l^m(\hat{\vec A_c})\int\sum_{t,u,v}
G_x^{A_{\mu}}(\tilde n,\tilde l,\tilde m,n,l,m,t,u,v)
\Lambda(\gamma,\vec r-\vec P,t,u,v)
\bigg[A(\vec r-\vec A_c)-\sum_{\vec n}^{pen}
\frac{1}{|\vec r-\vec A_c-\vec n|}\bigg]{\rm d^3r}
\end{eqnarray}

In similar way, the partial derivative with respect to center 
$\vec A_{\nu}$ is computed. 
Finally, the partial derivatives with respect to the 
centers $\vec A_c$ are obtained from translational invariance.

\subsubsection{Spheropole}

This term arises because the charge distribution is approximated by
a model charge distribution in the long range\cite{VicCoulomb}:

\begin{eqnarray}
\Phi^{ew}(\rho_c;\vec r)=
\Phi^{ew}(\rho_c^{model}; \vec r)+
\Phi^{ew}(\rho_c \! -\! \rho_c^{model};\vec r)=
\Phi^{ew}(\rho_c^{model}; \vec r)+\Phi^{coul}(\rho_c\!-\!
\rho_c^{model};\vec r)
+Q_c
\end{eqnarray}

The calculation of the 
Coulomb potential 
$\Phi^{coul}(\rho_c\!-\! \rho_c^{model};\vec r)$
is restricted to contributions from
those charges inside the penetration depth $pen$.
The use of the Coulomb potential 
$\Phi^{coul}(\rho_c\!-\!\rho_c^{model};\vec r)$ 
instead of the Ewald potential $\Phi^{ew}(\rho_c-\rho_c^{model};\vec r)$
is correct, if $\rho_c-\rho_c^{model}$ is of zero charge, dipole, 
quadrupole and spherical second moment\cite{Harris}. However, this 
condition leads to
a correction in the three-dimensional case\cite{Euwema,Harris,VicCoulomb}:
although the difference 
$\rho_c-\rho_c^{model}$ has zero charge, dipole and
quadrupole moment, it has in general a non-zero spherical second moment 
$Q_c$. Therefore, the potential must be shifted by $Q$ defined as:

\begin{eqnarray}
Q=\sum_{c}Q_c=
\sum_{c}\frac{2\pi}{3V}\int(\rho_c(\vec r)-\rho_c^{\rm model}(\vec r))
|\vec r|^2 
{\rm d^3r}
\end{eqnarray}

Three types of contributions are obtained \cite{VicCoulomb}:
zero, first and second order HGTFs. They have to be combined with
the corresponding $E$-coefficient. For the zeroth order, a contribution
of 

$E(\tilde n,\tilde l,\tilde m,n,l,m,0,0,0)
\left(\frac{3}{2\gamma}+\left(\vec A_{\mu}-\vec P\right)^2\right)$ 

is computed. The
derivative is therefore

\begin{eqnarray}
\frac{\partial}{\partial A_{\mu,x}}
\left(E(\tilde n,\tilde l,\tilde m,n,l,m,0,0,0)
\left(\frac{3}{2\gamma}+(\vec A_{\mu}-\vec P)^2\right)\right)
\end{eqnarray}

To obtain the derivative 
$\frac{\partial}{\partial A_{\mu,x}}E(\tilde n,\tilde l,\tilde m,n,l,m,0,0,0)$,
we use the identity
 
\begin{eqnarray} & &
\frac{\partial}{\partial A_{\mu,x}} 
\left(\sum_{t,u,v}E(\tilde n,\tilde l,\tilde m,n,l,m,t,u,v)
\Lambda(\gamma,\vec r-\vec P,t,u,v)\right)= \nonumber \\ & &
\sum_{t,u,v}E(\tilde n,\tilde l,\tilde m,n,l,m,t,u,v) 
\frac{\tilde\alpha}{\gamma}
\Lambda(\gamma,\vec r-\vec P,t+1,u,v)+ \nonumber \\ & & \sum_{t,u,v}
\Lambda(\gamma,\vec r-\vec P,t,u,v)
\frac{\partial}{\partial A_{\mu,x}}E(\tilde n,\tilde l,\tilde m,n,l,m,t,u,v)=  
\nonumber  \\ & &
\sum_{t,u,v} G_x^{A_{\mu}}(\tilde n,\tilde l,\tilde m,n,l,m,t,u,v)
\Lambda(\gamma,\vec r-\vec P,t,u,v)
\end{eqnarray}

which gives

\begin{eqnarray}
\frac{\partial}{\partial A_{\mu,x}} E(\tilde n,\tilde l,\tilde m,n,l,m,t,u,v)
=G_x^{A_{\mu}}(\tilde n,\tilde l,\tilde m,n,l,m,t,u,v)
-\frac{\tilde\alpha}{\gamma}
E(\tilde n,\tilde l,\tilde m,n,l,m,t-1,u,v) 
\end{eqnarray}

A similar operation is
necessary for the components with 
$E(\tilde n,\tilde l,\tilde m,n,l,m,1,0,0)$, 
$E(\tilde n,\tilde l,\tilde m,n,l,m,0,1,0)$ and 
$E(\tilde n,\tilde l,\tilde m,n,l,m,0,0,1)$ (first order HGTFs) 
which are multiplied
with prefactors $2(P_x-A_{\mu,x})$, $2(P_y-A_{\mu,y})$ and
$2(P_z-A_{\mu,z})$, respectively.
Finally, derivatives of the products of
$E(\tilde n,\tilde l,\tilde m,n,l,m,2,0,0)$, 
$E(\tilde n,\tilde l,\tilde m,n,l,m,0,2,0)$ and
$E(\tilde n,\tilde l,\tilde m,n,l,m,0,0,2)$ (second order HGTFs) 
with 2 are required.

\subsection{Bielectronic integrals}

We define a bielectronic integral as

\begin{eqnarray} & &
B_{\mu\vec 0\nu\vec g\tau\vec n \sigma\vec n+\vec h}= \nonumber \\ & &
\!\int\!\frac {\phi_{\mu}(\alpha_1,\vec r\!-\!\vec A_{\mu},n_1,l_1,m_1)
\phi_{\nu}(\alpha_2,\vec r\!-\!\vec A_{\nu}\!-\!\vec g,n_2,l_2,m_2)}
{|\vec r-\vec r \: '|} \nonumber \\ & &
\phi_{\tau}(\alpha_3,\!\vec r\:'\!-\!\!\vec A_{\tau}\!-\!\vec n,n_3,l_3,m_3
)\phi_{\sigma}
(\alpha_4,\vec r\:'\!\!-\!\!\vec A_{\sigma}\!-\!\vec n\!-\!\vec h,n_4,l_4,m_4)
 {\rm d^3r \: d^3r'}
= \nonumber \\ & &
\sum_{t,u,v}E(n_1,l_1,m_1,n_2,l_2,m_2,t,u,v)
 \sum_{t',u',v'}E (n_3,l_3,m_3,n_4,l_4,m_4,t',u',v')
[t,u,v|\frac{1}{|\vec r-\vec r \: '|}|t',u',v'] 
\end{eqnarray}

The expression $[t,u,v|\frac{1}{|\vec r-\vec r \: '|}|t',u',v']$ is 
defined as \cite{McMurchieDavidson,VicNATO}

\begin{eqnarray} & & 
[t,u,v|\frac{1}{|\vec r-\vec r \: '|}|t',u',v']= \nonumber \\ & & 
\int\int \Lambda(\gamma,\vec r-\vec P,t,u,v)\frac{1}{|\vec r-\vec r \: '|}
\Lambda(\gamma\:',\vec r\:'-\vec P\:',t,'u',v') {\rm d^3r\: d^3r\:'}
\end{eqnarray}

The partial derivative with respect to  $A_{\mu,x}$ is obtained as

\begin{eqnarray} & &
\frac{\partial}{\partial A_{\mu,x}} 
B_{\mu\vec 0\nu\vec g\tau\vec n \sigma\vec n+\vec h}= \nonumber \\ & &
\sum_{t,u,v}G_x^{A_{\mu}}(n_1,l_1,m_1,n_2,l_2,m_2,t,u,v)
 \sum_{t',u',v'}E (n_3,l_3,m_3,n_4,l_4,m_4,t',u',v')
[t,u,v|\frac{1}{|\vec r-\vec r \: '|}|t',u',v']
\end{eqnarray}

Similarly, gradients with respect to the other centers are obtained. One
of the gradients can be obtained by translational invariance if the other
three gradients have been computed.

In the context of periodic systems, it is necessary to perform summations
over the lattice vectors $\vec g, \vec h, \vec n$. We define a Coulomb
integral as follows

\begin{eqnarray} & &
C_{\mu\vec 0\nu\vec g\tau\vec 0\sigma \vec h}
= 
\sum_{\vec n}^{pen}B_{\mu\vec 0\nu\vec g\tau\vec n \sigma\vec n+\vec h}
\end{eqnarray}

Similarly, we define an exchange integral as follows:

\begin{eqnarray}
& &
X_{\mu\vec 0\nu\vec g\tau\vec 0\sigma \vec h}
= 
\sum_{\vec n}
B_{\mu\vec 0\tau\vec n \nu\vec g\sigma\vec n+\vec h} 
\end{eqnarray}

\section{Total energy and gradient}

\label{Calculationofderivatives}

\subsection{Total energy}

The correct summation of the Coulomb energy is the most severe problem
of the total energy calculation. The individual contributions to the
Coulomb energy, such as
for example the nuclear-nuclear interaction, are divergent for periodic
systems. Thus, a scheme based on the Ewald method is used to 
sum the individual contributions \cite{VicCoulomb}.
The total energy is then expressed as the sum of kinetic energy $E^{kin}$,
the Ewald energies of the nuclear-nuclear repulsion $E^{NN}$, 
nuclear-electron attraction $E^{coul-nuc}$, electron-electron repulsion
$E^{coul-el}$, and finally the exchange energy $E^{exch-el}$.

\begin{eqnarray} & & 
E^{\rm total}=E^{\rm kinetic}+E^{\rm NN}+E^{\rm coul-nuc}+E^{\rm coul-el}
+E^{\rm exch-el}=\nonumber \\ & & 
=\sum_{\vec g,\mu,\nu}P_{\nu\vec g\mu\vec 0}T_{\mu\vec 0\nu\vec g}+E^{\rm NN}
\nonumber \\ & & 
-\sum_{\vec g,\mu,\nu} 
P_{\nu\vec g\mu\vec 0}\sum_{a}Z_a\int
\phi_{\mu}(\tilde\alpha,\vec r-\vec A_{\mu},\tilde n,\tilde l,\tilde m)
\phi_{\nu}(\alpha,\vec r - \vec A_{\nu}-\vec g,n,l,m)
A(\vec r-\vec A_{a}){\rm d^3r}
\nonumber \\ & & 
+\frac{1}{2}\sum_{\vec g,\mu,\nu} P_{\nu\vec g\mu\vec 0}
\bigg(-QS_{\mu\vec 0\nu\vec g}+\sum_{\vec h,\tau,\sigma}
P_{\sigma\vec h\tau\vec 0}
C_{\mu\vec 0\nu\vec g \tau\vec 0\sigma\vec h}
-\sum_c \sum_{l=0}^{L}\sum_{m=-l}^{l}\eta_l^m(\rho_c;\vec A_c)
M_{l\mu\vec 0\nu\vec gc}^m\bigg)\nonumber \\ & & 
-\frac{1}{2}\sum_{\vec g,\mu,\nu}P^{\uparrow}_{\nu\vec g\mu\vec 0}
\sum_{\vec h,\tau,\sigma}P^{\uparrow}_{\sigma\vec h\tau\vec 0}
X_{\mu\vec 0\nu\vec g\tau\vec 0\sigma\vec h}
-\frac{1}{2}\sum_{\vec g,\mu,\nu}P^{\downarrow}_{\nu\vec g\mu\vec 0}
\sum_{\vec h,\tau,\sigma}P^{\downarrow}_{\sigma\vec h\tau\vec 0}
X_{\mu\vec 0\nu\vec g\tau\vec 0\sigma\vec h}
\end{eqnarray}

\subsection{Gradient of the total energy}

\label{Gradienttotenysection}

The force with respect to the position  of 
the nuclei can be calculated similarly to the molecular case
\cite{Bratoz,Pulay}. The derivatives of all the integrals are necessary,
and  the derivative of the density matrix is expressed with the help
of the energy-weighted density matrix. 
The full force is obtained as:

\begin{eqnarray}
\label{Forcegleichung} & &
\vec F_{A_i}=-\frac{\partial E^{\rm total}}{\partial \vec A_i}=\nonumber \\ & &
-\sum_{\vec g,\mu,\nu}P_{\nu\vec g\mu\vec 0}\frac{\partial 
T_{\mu\vec 0\nu\vec g}}
{\partial \vec A_i}
-\frac{\partial E^{\rm NN}}{\partial \vec A_i}\nonumber \\ & &
+\sum_{\vec g,\mu,\nu} P_{\nu\vec g\mu\vec 0}\sum_{a}Z_a
\frac{\partial}{\partial \vec A_i} \bigg[ \int
\phi_{\mu}(\alpha_2,\vec r-\vec A_{\mu},n_2,l_2,m_2)
\phi_{\nu}(\alpha_1,\vec r - \vec A_{\nu}-\vec g,n_1,l_1,m_1)
A(\vec r-\vec A_{a}){\rm d^3r}\bigg]
\nonumber \\ & &
-\frac{1}{2}\sum_{\vec g,\mu,\nu} P_{\nu\vec g\mu\vec 0}
\bigg\{-S_{\mu\vec 0\nu\vec g}
\frac{2\pi}{3V}
\sum_{c}
\sum_{\vec h,\sigma,\tau \in c}P_{\sigma \vec h \tau \vec 0}\nonumber \\ & &
\frac{\partial}{\partial \vec A_i} \int\bigg[
-\phi_{\tau}(\alpha_2,\vec r-\vec A_{\tau},n_2,l_2,m_2)
\phi_{\sigma}(\alpha_1, \vec r-\vec A_{\sigma}-\vec h,n_1,l_1,m_1)
\nonumber \\ & &
+\sum_{l=0}^L \sum_{m=-l}^l\int 
\phi_{\tau}(\alpha_2,\vec r \: '-\vec A{_\tau},n_2,l_2,m_2)
\phi_{\sigma}
(\alpha_1, \vec r \: '-\vec A_{\sigma}-\vec h,n_1,l_1,m_1)
X_l^m(\vec r \: '-\vec A_c){\rm d^3r'}
\delta_l^m(\vec A_c,\vec r)\bigg]r^2 {\rm d^3r} \nonumber\\ & &
+\sum_{\tau,\sigma}P_{\sigma\vec h\tau\vec 0}
\frac{\partial C_{\mu\vec 0\nu\vec g \tau\vec 0\sigma\vec h}}
{\partial \vec A_i}\nonumber \\ & &
+\sum_c \sum_{l=0}^{L}\sum_{m=-l}^{l}\sum_{\vec h,\tau \in  c, \sigma}
P_{\sigma\vec h \tau \vec 0} \nonumber \\ & & 
\frac{\partial}{\partial \vec A_i}\bigg[\int
\phi_{\tau}(\alpha_2,\vec r-\vec A_{\tau},n_2,l_2,m_2)
\phi_{\sigma}(\alpha_1,\vec r - \vec A_{\sigma}-\vec h,n_1,l_1,m_1)
X_l^m(\vec r-\vec {A_c}){\rm d^3r} \
 M_{l\mu\vec 0\nu\vec g c}^m\bigg]\bigg\}
\nonumber \\ & &
+\frac{1}{2}\sum_{\vec g,\mu,\nu} P^{\uparrow}_{\nu\vec g\mu\vec 0}
\sum_{\vec h,\tau,\sigma}P^{\uparrow}_{\sigma\vec h\tau\vec 0}
\frac{\partial X_{\mu\vec 0\nu\vec g\tau\vec 0\sigma\vec h}}
{\partial \vec A_i} 
+\frac{1}{2}\sum_{\vec g,\mu,\nu} P^{\downarrow}_{\nu\vec g\mu\vec 0}
\sum_{\vec h,\tau,\sigma}P^{\downarrow}_{\sigma\vec h\tau\vec 0}
\frac{\partial X_{\mu\vec 0\nu\vec g\tau\vec 0\sigma\vec h}}
{\partial \vec A_i} 
\nonumber \\ & &
+\sum_{\vec g,\mu,\nu}
\frac{\partial S_{\mu\vec 0\nu\vec g}}{\partial \vec A_i}  \int_{BZ} 
 \exp({\rm i}\vec k\vec g)\sum_j 
\left\{ \right.
a^{\uparrow}_{\nu j}(\vec k)a^{*\uparrow}_{\mu j}(\vec k)
(\epsilon^{\uparrow}_j(\vec k)+Q)
\Theta(\epsilon_F-\epsilon^{\uparrow}_j(\vec k)-Q) \nonumber \\ & & +
a^{\downarrow}_{\nu j}(\vec k)a^{*\downarrow}_{\mu j}(\vec k)
(\epsilon^{\downarrow}_j(\vec k)+Q)
\Theta(\epsilon_F-\epsilon^{\downarrow}_j(\vec k)-Q) \left. \right\}
{\rm d^3k}
\end{eqnarray}

The last addend is the energy weighted density matrix; the integral
is over the first Brillouin zone.

\section{Structure of the gradient code}
\label{Codestructure}
\begin{figure}
\caption{The present structure of the gradient code. The left column describes
the purpose of the routines, the middle column gives the names
of the corresponding 
routines, and the right column gives the name of the routines
in the energy code. One arrow indicates that the routine is a subroutine,
two arrows indicate that it is a subroutine called from a subroutine.}
\centerline{\psfig{figure=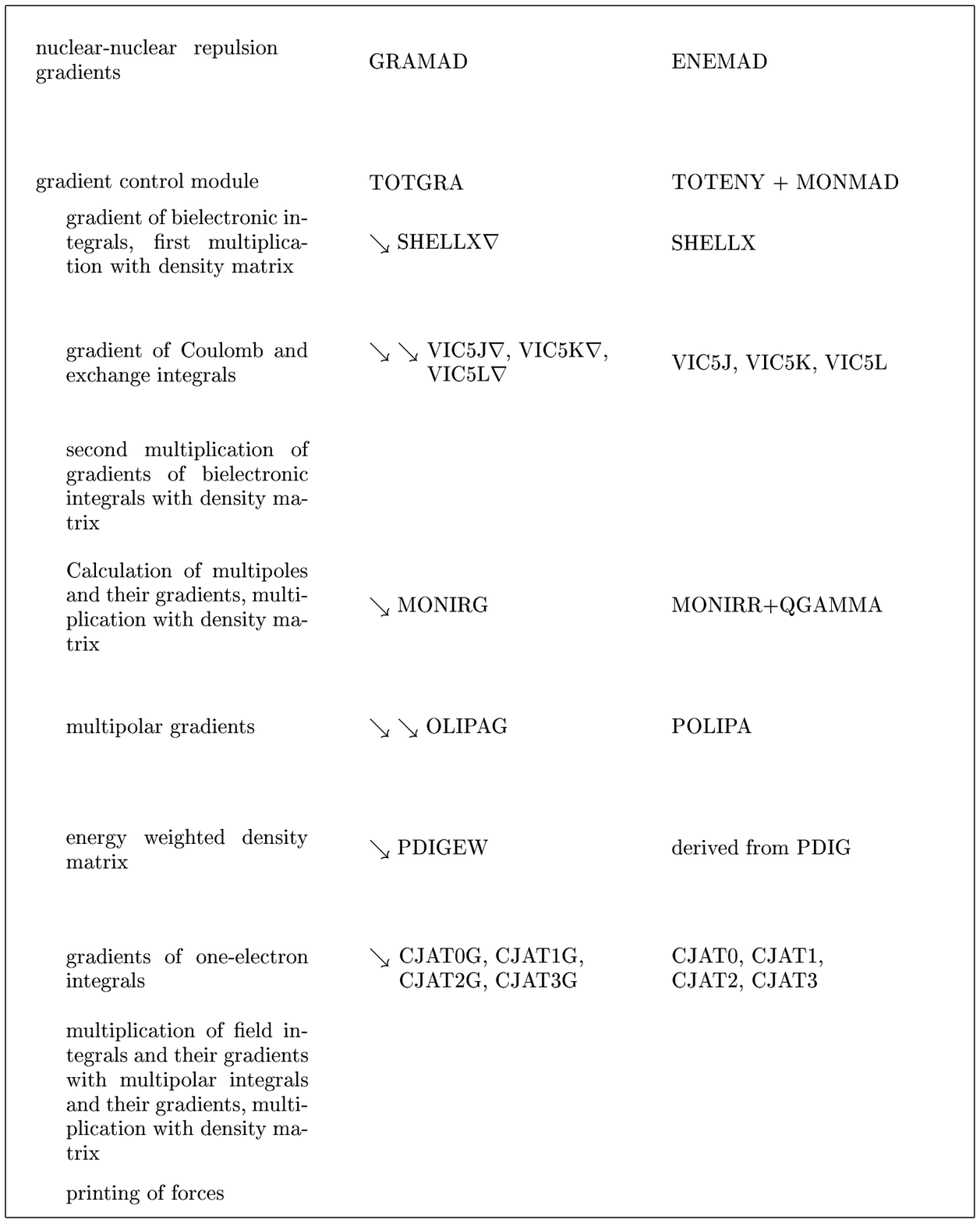,width=18cm,angle=0}}
\label{forceCPCfigure}
\end{figure}

The present structure of the gradient code is indicated in figure 
\ref{forceCPCfigure}. The first step is to compute the gradient of
the Ewald energy of the  nuclei 
in subroutine GRAMAD (the Ewald energy is computed in ENEMAD). The control
module TOTGRA then first calls routines to compute the gradient of the
bielectronic integrals (labeled with SHELLX$\nabla$ as these routines
will change their structure). The subroutine  SHELLX$\nabla$ calls
subroutines which explicitly compute the derivatives of Coulomb and
exchange integrals, and multiplies the gradients a first time with
the density matrix. Back in TOTGRA again, the second multiplication with
the density matrix is performed. The next step is to compute the derivatives
of the multipoles (MONIRG) and to compute the energy weighted density
matrix (PDIGEW). Then, the gradients of the one-electron integrals
are computed (CJAT0G, CJAT1G, CJAT2G or CJAT3G, depending on the dimension).
The field integrals and their gradients are now multiplied with the
multipolar integrals and their gradients, and a multiplication
with the density matrix is performed. This concludes the calculation
of the gradients. 

The structure has been simplified to focus on
the most important parts. 
In addition, as already mentioned, the code will undergo changes during
the optimization process so that a too detailed description seems to be 
unadvised.

\onecolumn
\newpage
\newpage

\section{Examples}
\label{Examplesection}

In tables \ref{NiO1D}, \ref{LiF2D} and \ref{NiO3D}, 
we give examples of the accuracy
of the gradients. First, in table \ref{NiO1D}, a chain of NiO molecules
is considered, with ferromagnetic  ordering (all the Ni spins up) and
with antiferromagnetic ordering (nearest Ni spins are antiparallel).
The oxygen atoms are moved by 0.01 \AA \ from their equilibrium positions
which results in a non-vanishing force. The agreement between
numerical and analytical gradient is better than 0.0001 $\frac{E_h}{a_0}$.
As we discussed in our first article \cite{IJQC}, the agreement can
be improved by using stricter "ITOL"-parameters (these are parameters
which control the accuracy of the evaluation of the 
integrals\cite{Manual}). Indeed, 
when increasing these parameters, the agreement further improves up
to an error of less than $10^{-5} \frac{E_h}{a_0}$.

In table \ref{LiF2D}, a LiF layer with a lattice constant of 5 \AA \
is considered with one atom being displaced from its equilibrium position. 
The forces agree to $2 \times 10^{-5} \frac{E_h}{a_0} $ when default
ITOL parameters (6, 6, 6, 6, 12) are used.

Finally, in table \ref{NiO3D}, a three-dimensional, ferromagnetically
polarized NiO solid is considered. When displacing the oxygen ions,
the forces agree to better than $2 \times 10^{-5} \frac{E_h}{a_0} $.

As a whole, the accuracy is certainly very high and can further
be improved by applying stricter cutoff (ITOL) parameters.

\section{Conclusion}

In this article, we described the implementation of analytical gradients
in the code CRYSTAL. In its present form, the code is capable of computing
highly accurate Hartree-Fock gradients for systems with 0,1,2 and 
3-dimensional periodicity. Both closed-shell restricted Hartree-Fock
as well as unrestricted Hartree-Fock 
calculations can be performed.

A first step of improving the efficiency of the code has been
completed with the coding of gradients for the bipolar expansion, and
a further enhancement of  the efficiency will be one of the future directions.
Of highest importance is the implementation of symmetry which will lead to
high saving factors \cite{Dovesi1986}.
Other targets are the implementation of gradients with respect to
the lattice vector, an extension to metallic 
systems\cite{Kertesz}, and the implementation
of density functional gradients.

\section{Acknowledgments}
The author would like to thank CCP3 and Prof. N. M. Harrison for their 
interest and support of this work (EPSRC grant GR/K90661),
Mr. V. R. Saunders for many helpful discussions, and Prof. R. Dovesi and the 
Turin group for helpful discussions and hospitality.

\newpage
\onecolumn

\begin{table}
\begin{center}
\caption{Ferromagnetic (FM) and antiferromagnetic (AFM) NiO chain (i.e. a chain
with alternating nickel and oxygen atoms). The distance
between two oxygen atoms is chosen as 5 \AA. The force is
computed numerically and analytically with the oxygen atoms
being displaced. A $[5s4p2d]$ basis set was used for nickel, and a
$[4s3p]$ basis set for oxygen.}
\label{NiO1D}
\begin{tabular}{ccccc} 
magnetic & ITOL &
displacement  & analytical derivative 
 & numerical derivative \\
ordering &  parameter &  of oxygen &  (x-component) & (x-component) \\
& & in \AA & $E_h/a_0$ & $E_h/a_0$\\
FM & 6 6 6 6 12 & 0.01 & 0.001274 & 0.001188 \\
FM & 8 8 8 8 14 & 0.01 & 0.001246 & 0.001249 \\
AFM & 6 6 6 6 12 & 0.01 & 0.001276 & 0.001191 \\
AFM & 8 8 8 8 14 & 0.01 & 0.001250 & 0.001252 \\
\end{tabular}
\end{center}
\end{table}

\begin{table}
\begin{center}
\caption{Forces on the atoms of a  LiF  layer when one of the atoms
is displaced from its equilibrium position. 
A $[4s3p]$ basis set was used for the fluorine atom and a $[2s1p]$
basis set for the lithium atom.
Default ITOL parameters were 
used.}
\label{LiF2D}
\begin{tabular}{ccc} 
atom & analytical derivative  & numerical derivative \\
& (x-component) & (x-component) \\
F at (0.5 \AA, 0 \AA, 0 \AA) &   0.001379 &  0.001400 \\
Li at (2.5 \AA, 0 \AA, 0 \AA) &  -0.020731 & -0.020726 \\
F at (2.5 \AA, 2.5 \AA, 0 \AA) &  0.010384 &  0.010376 \\
Li at (0 \AA, 2.5 \AA, 0 \AA) &   0.008969 &  0.008950 \\
\end{tabular}
\end{center}
\end{table}

\begin{table}
\begin{center}
\caption{\label{NiO3D} Ferromagnetic NiO in an fcc structure  
at a lattice constant of 4.2654 \AA. We compare
numerical and analytical derivatives when moving the oxygen ion parallel
to the x-direction. Default ITOL parameters were used, the
basis sets are the same as in table \ref{NiO1D}.}
\begin{tabular}{ccc} 
displacement of oxygen  & analytical derivative 
 & numerical derivative \\
& (x-component) & (x-component) \\
in \AA & $E_h/a_0$ & $E_h/a_0$\\
0.01 & 0.001499 & 0.001485 \\
0.02 & 0.002939 & 0.002925\\ 
0.03 & 0.004387 & 0.004378 \\
0.04 & 0.005857 & 0.005847 \\
0.05 & 0.007352 & 0.007346 \\
\end{tabular}
\end{center}
\end{table}

\end{document}